# Plasticity as Spontaneous Breaking of Symmetry

V. Kobelev[1]

**Abstract**
The Article demonstrates the spontaneous symmetry breaking of isotropic homogeneous elastic medium in form of transition from Euclidean to Riemann-Cartan internal geometry of medium. The deformation of elastic medium without defects is based on Euclidean geometry in three dimensional space. The deformation of elastic medium with defects is based on Riemann-Cartan geometry and is interpreted in this Article, as different phase state. In this article, the expression for the free energy leading is equal to a volume integral of the scalar function (the Lagrangian) that depends on metric and Ricci tensors only. In the linear elastic isotropic case the elastic potential is a quadratic function of the first and second invariants of strain and warp tensors with two Lame, two mixed and two bending constants. The conditions of stability of media are derived using expressions of free energy in two alternative phases. Initially the defect-free phase is stable and remains stable, when the strain is considerably small. With the increasing strain the stability conditions could be violated. If the strain in material attains the critical value, the pure geometric instability occurs. In other words, the medium undergoes the spontaneous symmetry breaking in form of emerging topological defects and transition to plastic flow.

[1] University of Siegen, Paul-Bonatz-Str. 9-11, D-57076, Siegen, Germany, kobelev@imr.mb.uni-siegen.de

**Lead Paragraph**

The aim of this study is the interpretation of transition from elastic state of material to plastic flow as the phase transition, which occurs as the spontaneous symmetry breaking. The isotropic material possess initially high grade of symmetry, which is characterized by Euclidean geometry and zero Riemann curvature and torsion. With the increasing elastic strain this symmetry spontaneously breaks and the state with non-vanishing curvature and torsion turns out to be more energetically preferable. This transition form internal Euclidean geometry of material to Riemann-Cartan geometry is treated as the phase transition. In the new phase the torsion could be physically interpreted as defects and the occurrence of defects results in plastic flow. In the adopted variational approach, the Lagrangian represent the free energy in solid with defects could be represented as the sum of three terms. One term describes the elastic energy of stress resultant tensor and the second the elastic energy of stress couple tensor. The third term is the mixed term, which contains the products of stress resultant and couple tensor. The Lagrangian is rewritten with the deformations and warp tensors. The positive-definiteness of Lagrangian expresses the stability of material. The conditions of stability are derived for the medium with defects in several cases analytically.

## 1 The deformation of elastic medium

For the beginning, consider infinite three dimensional elastic media. The media in the absence of defects is invariant under translations and rotations in reference coordinate system. The initial, or reference, configuration $s$ may be specified by means of a system of orthogonal curvilinear coordinates $q_i$. We shall suppose that this system moves continuously with body from the state $s$ to the state $S$ and so may be used as a system of curvilinear coordinates in $S$: $y^i = y^i(q^j)$, $x^i = x^i(q^i)$. The expressions for the squares of the line elements in $s$ and $S$ respectively are:

(1) $ds^2 = g_{ij} dq^i dq^j$, $dS^2 = G_{ij} dq^i dq^j$.

The state of strain of the elastic body may be characterized by the

(2) $dS^2 - ds^2 = 2\varepsilon_{ij} dq^i dq^j$.

The quantities $\varepsilon_{ij}$ are the components of symmetric strain tensor of the elastic body

(3) $2\varepsilon_{ij} = G_{ij} - g_{ij}$.

Well known, that the strain tensor characterizes the change of length and angles between line elements during the deformation of the body from the original configuration to its deformed configuration. In the absence of defects the (1) defines the Christoffel's symbols and the corresponding curvature tensors

(4) $\tilde{\Gamma}_{mnr} = \frac{1}{2}(\partial_m G_{nr} + \partial_n G_{mr} - \partial_r G_{mn})$, $\tilde{\gamma}_{mnr} = \frac{1}{2}(\partial_m g_{nr} + \partial_n g_{mr} - \partial_r g_{mn})$

(5) $\tilde{R}_{nmlk} = 2(\partial_n \tilde{\Gamma}_{mlk} - G^{pq} \tilde{\Gamma}_{nkq} \tilde{\Gamma}_{mlp})_{[nm]}$, $\tilde{r}_{nmlk} = 2(\partial_n \tilde{\gamma}_{mlk} - g^{pq} \tilde{\gamma}_{nkq} \tilde{\gamma}_{mlp})_{[nm]}$

These tensors in the absence of defects equal identically zero

$\mathbf{R} = 0, \mathbf{r} = 0$

due to the nonexistence of the curvature of the Euclidean space. For the same reason the torsion tensor of the deformed medium vanishes. Remarkable, that disappearance of curvature in the linear approximation is equivalent to the Saint-Venant's compatibility conditions. Thus, if the deformation occurs in the elastic body without defects, the torsion and curvature are zero in both original and deformed configurations.

The internal geometry of elastic medium with geometrical defects is entirely different and can not be considered as Euclidean, even being embedded in Euclidean three dimensional space. The deformation of elastic medium with geometrical defects is based on Riemann-Cartan geometry. In Riemann-Cartan geometry, the affine connection in reference configuration $s$ is defined uniquely by the tensors of metric $\mathbf{g}$, torsion $\mathbf{t}$ and nonmetricity $\mathbf{q}$:

(6) $\gamma_{mnr} = \frac{1}{2}(\partial_m g_{nr} + \partial_n g_{mr} - \partial_r g_{mn}) + \frac{1}{2}(t_{mnr} - t_{nmr} + t_{rmn}) + \frac{1}{2}(q_{mnr} + q_{nmr} - q_{rmn})$.

The affine connection in deformed configuration $S$ is defined by the tensors of metric $\mathbf{G}$, torsion $\mathbf{T}$ and nonmetricity $\mathbf{Q}$:

(7) $\Gamma_{mnr} = \frac{1}{2}(\partial_m G_{nr} + \partial_n G_{mr} - \partial_r G_{mn}) + \frac{1}{2}(T_{mnr} - T_{nmr} + T_{rmn}) + \frac{1}{2}(Q_{mnr} + Q_{nmr} - Q_{rmn})$.

The number of independent components of connection $\Gamma_{mn}^{\ s}$ in $n$ dimensions is $n^3$. The number of independent components of torsion $T_{mn}^{\ s}$ in $n$ dimensions is $(n-1)n^2/2$. The nonmetricity tensor possesses $(n+1)n^2/2$ independent components. Physically the nonmetricity can be associated with extra-matter [1].

Hereafter we consider only the metrical connections, such that the nonmetricity in both configurations $\mathbf{q}$ and $\mathbf{Q}$ vanish. In this case, the covariant derivatives of the metric are identically equal to zero

(8) $\nabla_m g_{nr} \equiv \partial_m g_{nr} - \gamma_{mn}^{\ s} g_{sr} - \gamma_{mr}^{\ s} g_{nr} = 0$, $\nabla_m G_{nr} \equiv \partial_m G_{nr} - \Gamma_{mn}^{\ s} G_{sr} - \Gamma_{mr}^{\ s} G_{nr} = 0$

The curvature tensors are defined via

(9) $r_{nmlk} = 2(\partial_n \gamma_{mlk} - g^{pq} \gamma_{nkq} \gamma_{mlp})_{[nm]}$, $R_{nmlk} = 2(\partial_n \Gamma_{mlk} - G^{pq} \Gamma_{nkq} \Gamma_{mlp})_{[nm]}$

Contraction of the curvature tensor in the two indices yields the Ricci tensors of the reference and deformed configurations

(10) $r^{ij} = \frac{1}{4}\epsilon^{jmn}\epsilon_{ilk}\, r_{nmlk}$, $\qquad R^{ij} = \frac{1}{4}\epsilon^{jmn}\epsilon_{ilk}\, R_{nmlk}$

In three-dimensional space solely the Ricci tensor defines the curvature tensor

(11) $r_{ijkl} = \delta_{ik} r_{jl} - \delta_{il} r_{jk} - \delta_{jk} r_{il} + \delta_{jl} r_{ik} - \frac{1}{2}(\delta_{ik}\delta_{jl} - \delta_{il}\delta_{jk})r$,

(12) $r = g^{ij} r_{ij}$.

Consider now the locally elastic body that moves from its original configuration $s$ to the deformed configuration $S$. It is assumed, that the deformation of the body is accompanied both by the appearance of torsion and curvature, or, in other words, by the appearance of defects.

## 2  The Lagrangian of the elastic medium

For the estimation of free energy of medium with defects, we need the expression of Lagrangian associated with torsion. In this article, the expression for the free energy is equal to a volume integral of the scalar function (the Lagrangian) that depends on metric and Ricci tensors only.

The warp tensor $\boldsymbol{\kappa}$ characterizes the deviation of local curvature of medium during the deformation of the body from the original configuration to its deformed configuration:

(13)  $2\boldsymbol{\kappa} = \mathbf{R} - \mathbf{r}$.

The components of warp tensor could be expressed by the difference of components of Ricci tensors in original and deformed configurations

(14)  $2\kappa_{ij} = R_{ij} - r_{ij}$

In the static theory for the media with dislocations and disclinations the possible choice of the geometric Lagrangian yield the equations of equilibrium.

For the description of deformation for homogeneous isotropic body we require the invariants of tensors. The invariants of the metric tensor in states $s$ and $S$ are respectively $g_I, g_{II}, g_{III}$ and $G_I, G_{II}, G_{III}$. The values $r_I, r_{II}, r_{III}$ and $R_I, R_{II}, R_{III}$ are the invariants of the symmetric parts of Ricci tensor in state $s$ and in state $S$ correspondingly.

The warp tensor is symmetric with respect to the permutation of indices. In Riemann-Cartan geometry, this symmetry for a nonzero torsion tensor $T_{ijk}$ is not present in common case. The Ricci tensors could be spited in its symmetric $\hat{r}_{ij} = r_{(ij)}$, $\hat{R}_{ij} = R_{(ij)}$ and skew-symmetric parts $\bar{r}_{ij} = r_{[ij]}$, $\bar{R}_{ij} = R_{[ij]}$. The warp tensor splits in its symmetric $\hat{\kappa}_{ij} = \kappa_{(ij)}$ and skew-symmetric parts $\bar{\kappa}_{ij} = \kappa_{[ij]}$. The first and third invariants of skew-symmetric parts all vanish ($\bar{K}_I = \bar{\kappa}_i^{\ i} = 0$, $\bar{K}_{III} = \bar{\kappa}_i^{\ j}\bar{\kappa}_j^{\ k}\bar{\kappa}_k^{\ i} = 0$). The skew-symmetric parts lead to additional terms in the elastic energy, containing the terms, which depend on the second invariant of skew-symmetric part of Ricci tensor

$$\bar{r}_{II} = r_{[ij]}r^{[ji]},\ \bar{R}_{II} = R_{[ij]}R^{[ji]},\ \bar{K}_{II} = \bar{\kappa}_{ij}\bar{\kappa}^{ji}.$$

The elastic potential of homogeneous isotropic body is a function of metric tensor invariants

$$g_I, g_{II}, g_{III}, G_I, G_{II}, G_{III}$$

and Ricci tensor invariants

$$r_I, r_{II}, \bar{r}_{II}, r_{III}, R_I, R_{II}, \bar{R}_{II}, R_{III}.$$

The elastic potential per unit volume of the unstrained body reads

(15) $\quad L = W(G_I, g_I, G_{II}, g_{II}, G_{III}, g_{III}, R_I, r_I, R_{II}, r_{II}, \bar{R}_{II}, \bar{r}_{II}, R_{III}, r_{III}).$

The metric of the original configuration is arbitrary, such that only the change of metric and curvature during the deformation results in change of energy density. Therefore in solid mechanics the use of strain invariants $E_I, E_{II}, E_{III}$ instead of metric tensor invariants $g_I, g_{II}, g_{III}, G_I, G_{II}, G_{III}$ is more common. Similarly, we use warp invariants $K_I, K_{II}, \bar{K}_{II}, K_{III}$ instead of Ricci tensor invariants $r_I, r_{II}, \bar{r}_{II}, r_{III}, R_I, R_{II}, \bar{R}_{II}, R_{III}$.

One can use the invariants in the form

(16) $\quad E_I = \varepsilon_i^{\ i},\qquad E_{II} = \varepsilon_{ij}\,\varepsilon^{ij},\qquad E_{III} = \varepsilon_i^{\ j}\varepsilon_j^{\ k}\varepsilon_k^{\ i}$

(17) $\quad K_I = \hat{\kappa}_i^{\ i},\qquad K_{II} = \hat{\kappa}_{ij}\,\hat{\kappa}^{ij},\qquad K_{III} = \hat{\kappa}_i^{\ j}\hat{\kappa}_j^{\ k}\hat{\kappa}_k^{\ i}$

To obtain the explicit relation between strain and warp to stress and moments, we must determine from the experiments the function $W$. In its turn, this implies that we must make some physical assumptions about the nature of the bodies being deformed. In the elasticity theory, the elastic potential depends only on the three strain invariants and on scalar function of coordinates [2]. The variational methods in the medium with continuously distributed dislocations were developed in [3,4]. According to [5], the expression for the free energy is equal to a volume integral of the scalar function (the Lagrangian) that is quadratic in strain, torsion and curvature tensors. The expression for the free energy leading to the equilibrium equations is usually assumed to be equal to a volume integral of the scalar function (the Lagrangian) that is quadratic in strain and torsion [6, 7].

We shall draw our attention to bodies which are of constant density in the unstrained state and whose elastic potential depends only on the three strain invariants, three warp invariants and on scalar function of coordinates. The similar elastic potential is defined in a micropolar elasticity theory [8]. The most common expression for elastic potential for homogeneous isotropic elastic medium is assumed in the following form

(18) $\quad L = W(E_I, E_{II}, E_{III}, K_I, K_{II}, \bar{K}_{II}, K_{III}).$

For the linear theory of homogeneous anisotropic elastic medium the elastic potential must be quadratic in warp and strain, such that

(19) $\quad 2L = C^{ij}_{\ \ kl}\varepsilon_{ij}\varepsilon^{kl} + 2B^{ij}_{\ \ kl}\varepsilon_{ij}\kappa^{kl} + D^{ij}_{\ \ kl}\kappa_{ij}\kappa^{kl}.$

Here $\quad C^{ij}_{\ \ kl}\quad$ are the components of elastic constants tensor with the dimension of $Pa$,

$\qquad B^{ij}_{\ \ kl}\quad$ are the mixed elastic constants with the dimension of $N$,

$\qquad D^{ij}_{\ \ kl}\quad$ are the bending modules with the dimension of $Nm^2$.

Under the presuming of the hypothesis of central symmetry [9], the coefficients $B^{ij}_{\ \ kl}$ vanish.

In the linear elastic isotropic case the elastic potential is a quadratic function of the first and second invariants of strain and warp tensors

(20) $\quad 2L = \lambda\varepsilon^i_{\ i}\varepsilon^k_{\ k} + 2l\varepsilon^i_{\ i}\kappa^k_{\ k} + \Lambda\kappa^j_{\ i}\kappa^k_{\ k} + 2\mu\varepsilon_{ik}\varepsilon^{ik} + 4m\varepsilon_{ik}\kappa^{ik} + 2M\kappa_{ki}\kappa^{ki} + 2H\kappa_{ki}\kappa^{jk}.$

Here $\lambda, \mu$ $\qquad$ are Lame constants with the dimension of $Pa = Nm^{-2}$,

$\qquad$ l, m $\qquad$ are the mixed elastic constants with the dimension of $N$,

$\qquad \Lambda, M, H \qquad$ are the bending modules with the dimension of $Nm^2$.

The constants for the elastic potential possess a clear physical meaning. With the hypothesis of central symmetry the isotropic elastic potential yields

$$L = L_\varepsilon + L_\kappa,$$

(21) $\quad 2L_\varepsilon = \lambda \varepsilon^i{}_i \varepsilon^k{}_k + 2\mu \varepsilon_{ik} \varepsilon^{ik} \equiv \lambda E_I^2 + 2\mu E_{II},$

$$2L_\kappa = \Lambda \kappa^i{}_i \kappa^k{}_k + 2\mathrm{M} \kappa_{ki} \kappa^{ki} + 2\mathrm{H} \kappa_{ki} \kappa^{ik} \equiv \Lambda K_I^2 + 2\mathrm{M} K_{II} + 2\mathrm{H} \overline{K}_{II}.$$

This form of elastic potential uses only five constants. The differential equations of the medium could be obtained from this equation using the common method, outlined in [2]. The independent variables are the displacements and components of torsion and nonmetricity.

## 3  The physical meaning of torsion in the mechanics of elastic medium

In this section we reveal the known physical meaning of torsion and curvature in the mechanics of elastic medium with defects. Torsion **t** defines the tensor **α** by

(22) $\quad \alpha^{ij} = \in^{ikm} t^j{}_{km}, \qquad A^{ij} = \in^{ikm} T^j{}_{km}$

In the media with continuously distributed dislocations the tensors **α** and **A** can be identified with the dislocation tensors in the reference and deformed configurations.

The incompatibility tensors [10, 11] **θ** in the reference and **Θ** in the deformed configurations can be defined locally by the curvature tensors **r** and **R** of the forth rang

(23) $\quad \theta^{ij} = \dfrac{1}{2} \in^{imn} \in^{jkl} r_{klmn}, \qquad \Theta^{ij} = \dfrac{1}{2} \in^{imn} \in^{jkl} R_{klmn}.$

If there are continuously distributed disclinations in the media, the incompatibility tensor can be identified with the disclination tensor. With this formula the Ricci tensor could be expressed by the disclination tensor

(24) $\quad 2r^{ij} = \theta^{ij}, \qquad 2R^{ij} = \Theta^{ij}.$

The Bianchi identities read for dislocation tensors **α**, **A** and disclination tensors **θ**, **Θ** as [12]

(25) $\quad \theta^i{}_{j,i} = 0, \qquad \alpha^i{}_{m,i} = -\in_m{}^{ij} \theta_{ij}$

(26) $\quad \Theta^i{}_{j,i} = 0, \qquad A^i{}_{m,i} = -\in_m{}^{ij} \Theta_{ij}$

The Burgers **b**, **B** and Frank vectors **ω**, **Ω** are expressed for any closed path $\Gamma$ through the dislocation tensor **α**, **A** and disclination (incompatibility) tensor **θ**, **Θ** respectively as

(27) $\quad \oint_\Gamma \alpha_{ij} ds^j = b_i, \qquad \oint_\Gamma \theta_{ij} ds^j = \omega_i,$

(28) $\quad \oint_\Gamma A_{ij} ds^j = B_i, \qquad \oint_\Gamma \Theta_{ij} ds^j = \Omega_i.$

The media containing dislocations represents itself the Euclidean space. The displacement vector in the media in the presence of dislocations is due to dislocation lines not a smooth function. Dislocations in elastic medium could be represented by the dislocation line and the Burgers vector. The presence of dislocations in the medium with defects gives rise for the torsion tensor. The physical interpretation of the torsion tensor is the surface density of the Burgers vector.

Analogously, the curvature tensor defines the surface density of the Frank vector characterizing the disclinations in the spin structure. The physical interpretation of the curvature tensor is the surface density of the Frank vector.

The dislocation and disclination tensors may alter during the deformation. There exist four distinct cases. Firstly, both the dislocation and disclination tensors vanish in both states $s$ and $S$. The torsion and the curvature in both states vanish, and the deformation characterizes by the strain tensor **ε** only. In this case the medium behaves elastically during the deformation.

Secondly, the dislocation and disclination tensors vanish in the initial state $s$, but not identically zero in the state $S$. Both torsion and incompatibility vanish in the initial state and the geometry of the body in its initial state is Euclidean. The deformed state of the medium depends on the dislocation tensor **A** and disclination (incompatibility) tensor **Θ** in the deformed state $S$. The geometry of the body in its deformed state is non-Euclidean. If only dislocations appear during the deformation, the deformed geometry is Cartanian. If only disclinations emerge, the deformed geometry is Riemannian. If both dislocations and deformation appear, the deformed body is described by Riemann-Cartan geometry, which characterizes by tensors of metric **G**, Ricci tensor **R** and torsion **T**.

Thirdly, the dislocation and disclination tensors vanish in the deformed state $S$, but not identically zero in the initial state $s$. The deformed state of the medium depends on the dislocation tensor **α** and disclination (incompatibility) tensor **θ** in the initial state $s$.

Fourthly, the dislocation and disclination tensors non-vanish in the initial state $s$ and in the deformed state $S$. The deformed state of the medium depends on the dislocation tensor **α** and disclination (incompatibility) tensor **θ** in initial state $s$ and on the dislocation tensor **A** and disclination (incompatibility) tensor **Θ** deformed state $S$.

Thus, the geometry of both configurations is characterized by the covariant metric tensor **g** and Ricci tensor **r** in original configuration $s$ and the covariant metric tensor **G** and Ricci tensor **R** in deformed configuration $S$.

In the current study we do not prescribe torsion tensor as a priory given. On the contrary, we treat the torsion tensor as dynamical variable. In the initial state $s$ the torsion tensor vanishes. In the deformed state $S$ .the torsion tensor is generally saying, non-vanish. The non-vanishing of torsion tensor corresponds to emergence of plastic flow in the material.

The defects in the solids are of diverse nature and of different scale. The microscopic defects arise in a consequence of atomic misplacement in spatial lattices. The microscopic defects are the source of local warp. The hypothesis of defects, considered in this Article, is simple yet comprehensive and free of additional physical assumptions.

## 4 Stability conditions and spontaneous symmetry breaking in two-dimensional case

For plane deformation $n = 2$ we introduce the following notation for two remaining nonzero components of torsion $t^1 \equiv t^1_{12}, t^2 \equiv t^2_{12}$ in the initial state $s$ and $T^1 \equiv T^1_{12}, T^2 \equiv T^2_{12}$ in the deformed state $S$. The Ricci tensor in the deformed state $S$ reads

(29) $$\mathbf{R} = \frac{1}{2}\begin{bmatrix} \partial_2 T^1 & \partial_2 T^2 \\ -\partial_1 T^1 & -\partial_1 T^2 \end{bmatrix},$$

(30) $$\hat{\mathbf{R}} = \frac{1}{4}\begin{bmatrix} 2\partial_2 T^1 & \partial_2 T^2 - \partial_1 T^1 \\ \partial_2 T^2 - \partial_1 T^1 & -2\partial_1 T^2 \end{bmatrix},$$

(31) $$\overline{\mathbf{R}} = \frac{1}{4}\begin{bmatrix} 0 & \partial_2 T^2 + \partial_1 T^1 \\ -\partial_2 T^2 - \partial_1 T^1 & 0 \end{bmatrix}.$$

The similar expressions take place in the initial state $s$. The invariants of Ricci tensor **R** yield

(32) $$R_I = g^{ij} R_{ij} = \frac{1}{2}(\partial_2 T^1 - \partial_1 T^2),$$

(33) $$R_{II} = R_{(ij)} R^{(ij)} = \tfrac{1}{4}(\partial_2 T^1)^2 + \tfrac{1}{8}(\partial_2 T^2 - \partial_2 T^2)^2 + \tfrac{1}{4}(\partial_1 T^2)^2,$$

(34) $$\overline{R}_{II} = R_{[ij]} R^{[ji]} = -\tfrac{1}{8}(\partial_2 T^2)^2 - \tfrac{1}{8}(\partial_1 T^1)^2.$$

The part of Lagrangian (21) of the medium with central symmetry, which depends on Ricci tensor (29)..(34), in matrix form yields

(35) $$L_K = \tfrac{1}{2} \mathbf{v}^T \mathbf{M} \mathbf{v}.$$

Here **v** is the vector, depending only on torsion components and **M** is the matrix, depending only on elastic constants of material

$$(36) \quad \mathbf{v} = \begin{Vmatrix} \partial_2(T^1 - t^1) \\ \partial_1(T^2 - t^2) \\ \partial_1(T^1 - t^1) \\ \partial_2(T^2 - t^2) \end{Vmatrix}, \quad \mathbf{M} = \begin{Vmatrix} m_{11} & m_{12} & 0 & 0 \\ m_{12} & m_{22} & 0 & 0 \\ 0 & 0 & m_{33} & m_{34} \\ 0 & 0 & m_{34} & m_{44} \end{Vmatrix}.$$

The non-vanishing coefficients of matrix $\mathbf{M}$ are

$m_{11} = m_{22} = \Lambda + 2\mathrm{M}$, $\quad m_{12} = -\Lambda$, $\quad m_{34} = \mathrm{H}/2 - \mathrm{M}$, $\quad m_{33} = m_{44} = \mathrm{H}/2 + \mathrm{M}$.

The characteristic polynomial of matrix $\mathbf{M}$ reads

$$F_0(x) = |\mathbf{M} - \mathbf{E}x| = (2\mathrm{M} - x)^2 (2(\mathrm{M} + \Lambda) - x)(\mathrm{H} - x)$$

The eigenvalues of the matrix $\mathbf{M}$ are $\mathrm{H}, 2\mathrm{M}, 2\mathrm{M}, 2\mathrm{M} + 2\Lambda$. The condition for positive definiteness of energy (35) is equivalent to positivity of all eigenvalues of the matrix $\mathbf{M}$:

$\mathrm{H} > 0$, $\quad \mathrm{M} > 0$, $\quad \Lambda + \mathrm{M} > 0$.

This is the set of conditions on the elastic constants of the medium that must be satisfied if the medium is to maintain structural stability in its free state.

We precede with study the interaction between strain and wrap. This issue plays the central role for the development of material model. In the current model the physical defects of the material – dislocations and disclinations – are represented by geometrical object – wrap tensor, which depends upon torsion and curvature. In the absence of defects the wrap tensors vanishes. The application of load leads to the macroscopic deformation of material. For the elastic material, the macroscopic deformation of material is the change of metric. The curvature of the material in its reference and deformed states vanish, as the material is embedded in the Euclidean space. In the theory of the geometric description of defects, the inelastic behavior is represented by the appearance and change of torsion and curvature of material.

A simple model for inelastic behavior of material is based on the geometric description of defects and stability conditions of the medium with appearing defects. The condition of the emergence of defects is investigated with the expressions (21) for free energy. The medium with pre-strain, but without defects, is considered as the reference state $s$.

Firstly, in the absence of strain the wrap-dependent part of Lagrangian is positively defined $L_\kappa > 0$. This prohibits the spontaneous appearance of defects in the absence of sufficiently small applied strain also.

Secondly, if the strain in material exceeds certain level, the spontaneous creation of defects might provide less free energy density and be favorable from viewpoint of energy. The medium with the same pre-strain, but with newly borne topological defects, is considered as the deformed state $S$. The emerged defects contribute to the macroscopic deformation of material. The strain level, which is required for the appearance of defects, will be determined in this study immediately from the stability conditions of pre-strained material. If defects move and remain in material after load declines, this mechanism could lead to inelastic behavior of material, but this process is not a subject of the current study.

The metric tensors in the reference state $s$ and deformed state $S$ are taken in the form

$$(37) \quad \mathbf{g} = \mathbf{G} = \begin{bmatrix} 1 + 2\varepsilon_{11} \cdot \delta & 1 + 2\varepsilon_{12} \cdot \delta \\ 1 + 2\varepsilon_{12} \cdot \delta & 1 + 2\varepsilon_{22} \cdot \delta \end{bmatrix}.$$

Here $\varepsilon_{ij}$ are the components of the strain tensor $\boldsymbol{\varepsilon}$. The small parameter $\delta$ is required for the subsequent Taylor expansions. In this section, we restrict the consideration to two-dimensional case.

The components of torsion vanish $\gamma_1 \equiv \gamma_{112} = 0$, $\gamma_2 \equiv \gamma_{122} = 0$ in the initial state $s$. The Ricci tensor of the initial state $s$ vanishes as well. Physically this means, that the elastic medium in the initial state $s$ is pre-strained, but there are no dislocations or disclinations in this state. The medium behaves linear elastically.

The only two non-vanishing components of torsion in the deformed state $S$ are $T_1 \equiv \delta \cdot T_{112}$, $T_2 \equiv \delta \cdot T_{122}$. Physically this means, that the elastic medium in the deformed state $S$ is pre-strained, and the dislocations or disclinations could present. The decision, if the dislocations or disclinations really present depends on free energy in the deformed state. If the free energy density in the medium with dislocations and disclinations is lower, than the free energy in the medium without defects, the defects emerge. If the free energy density in the medium with dislocations and disclinations is higher, than the free energy in the medium without defects, the defects absent in deformed state $S$ as well. Note, that the energy quantity, which is necessary for the creation of defects is neglected in this study for clarity and could be easily accounted.

Non-vanishing connection components are

$$\Gamma_{111} = \delta \cdot \left(\frac{1}{2} Q_{111} + \partial_1 \varepsilon_{11}\right), \quad \Gamma_{112} = \delta \cdot \left(\frac{1}{2} Q_{211} + T_1 + 2\partial_1 \varepsilon_{12} - \partial_2 \varepsilon_{11}\right),$$

$$\Gamma_{121} = \delta \cdot \left(Q_{112} - \frac{1}{2} Q_{211} + \partial_2 \varepsilon_{11}\right), \quad \Gamma_{122} = \delta \cdot \left(\frac{1}{2} Q_{122} + T_2 + \partial_1 \varepsilon_{22}\right),$$

$$\Gamma_{222} = \delta \cdot \left(\frac{1}{2} Q_{222} + \partial_2 \varepsilon_{22}\right), \quad \Gamma_{221} = \delta \cdot \left(\frac{1}{2} Q_{122} - T_2 + 2\partial_2 \varepsilon_{12} - \partial_1 \varepsilon_{22}\right),$$

$$\Gamma_{212} = \delta \cdot \left(Q_{212} - \frac{1}{2} Q_{112} + \partial_1 \varepsilon_{22}\right), \quad \Gamma_{211} = \delta \cdot \left(\frac{1}{2} Q_{211} - T_1 + \partial_2 \varepsilon_{11}\right).$$

With these expressions we get the Ricci tensor of the deformed configuration $S$. In the subsequent study the nonmetricity is assumed to be identically zero.

For study of stability we need only two first terms of Taylor series for Ricci tensor

(38)   $\mathbf{R} = \mathbf{\Psi} \cdot \delta + (\mathbf{\Xi} + \overline{\mathbf{\Xi}}) \cdot \delta^2 + o(\delta^2)$.

In (38) $\mathbf{\Psi}$, $\mathbf{\Xi}$, $\overline{\mathbf{\Xi}}$ denote the tensors with the following components in the orthogonal Cartesian coordinate system:

$$\Psi_{11} = \partial_2 T_1, \quad \Psi_{12} = \partial_2 T_2, \quad \Psi_{21} = -\partial_1 T_1, \quad \Psi_{22} = -\partial_1 T_2,$$

$$\Xi_{11} = T_2 \cdot \partial_1 \varepsilon_{22} + T_1 \cdot \partial_2 \varepsilon_{11} - 2\partial_2 T_1 \cdot \varepsilon_{22} - 2\partial_1 T_1 \cdot \varepsilon_{12}$$

$$\Xi_{12} = -T_1 \cdot \partial_1 \varepsilon_{22} + T_2 \cdot \partial_2 \varepsilon_{22} + 2T_1 \cdot \partial_2 \varepsilon_{12} - 2\partial_2 T_2 \cdot \varepsilon_{22} - 2\partial_1 T_2 \cdot \varepsilon_{12}$$

$$\Xi_{21} = T_2 \cdot \partial_2 \varepsilon_{11} - T_1 \cdot \partial_1 \varepsilon_{11} - 2T_2 \cdot \partial_1 \varepsilon_{12} + 2\partial_1 T_1 \cdot \varepsilon_{11} + 2\partial_2 T_1 \cdot \varepsilon_{12}$$

$$\Xi_{22} = -T_1 \cdot \partial_2 \varepsilon_{11} - T_2 \cdot \partial_1 \varepsilon_{22} + 2\partial_1 T_2 \cdot \varepsilon_{11} + 2\partial_2 T_2 \cdot \varepsilon_{12}$$

$$\overline{\Xi}_{11} = (T_2)^2, \quad \overline{\Xi}_{12} = -T_1 T_2, \quad \overline{\Xi}_{21} = -T_1 T_2, \quad \overline{\Xi}_{22} = (T_1)^2.$$

The substitution of Ricci tensor (38) into Lagrangian (21) delivers the expression for elastic energy of defects in terms of torsion and pre-strain. The free energy is a nonlinear function of strain and torsion. The specific form of the free energy expansion is established by the analogous considerations as in the general theory of phase transitions of the second kind [13]. The emergence of torsion – and with torsion associated defects – is treated as the phase transition. The linear elastic material undergoes the phase transition and turns into phase of plastic deformation. This transition will be studied using the methods of stability of dynamical systems [14].

As discussed in the previous section, the strain-free medium is stable. With the increasing strain, however, the stability conditions could be violated. For definiteness, consider the case of homogeneous principal strains $\varepsilon_{11}$, $\varepsilon_{22}$. The principal strains do not depend on coordinates and $\varepsilon_{12} = 0$.

Our aim is to express the stability conditions in terms of applied principal strains. The part of free energy, which depends on strains, remains constant, as the strains in the initial and deformed states do not change. The part of free energy, which depends on warp, modifies due to the presence of defects. The warp-depending part of Lagrangian (21) in the deformed state reads

(39)   $L_\kappa = \frac{1}{2} \hat{\mathbf{v}}^T \hat{\mathbf{M}} \hat{\mathbf{v}}$

with

(40)   $\hat{\mathbf{v}} = \begin{Vmatrix} \partial_2 T^1 \\ \partial_1 T^2 \\ \partial_1 T^1 \\ \partial_2 T^2 \end{Vmatrix}, \quad \hat{\mathbf{M}} = \begin{Vmatrix} \hat{m}_{11} & \hat{m}_{12} & 0 & 0 \\ \hat{m}_{12} & \hat{m}_{22} & 0 & 0 \\ 0 & 0 & \hat{m}_{33} & \hat{m}_{34} \\ 0 & 0 & \hat{m}_{34} & \hat{m}_{44} \end{Vmatrix}.$

The components of matrix $\hat{\mathbf{M}}$ are

$$\hat{m}_{11} = m_{22} = (\Lambda + 2M)(1 - 4\varepsilon_{11} - 4\varepsilon_{22}),$$
$$\hat{m}_{12} = -\Lambda(1 - 4\varepsilon_{11} - 4\varepsilon_{22}),$$
$$\hat{m}_{34} = (H - 2M)(1/2 - 2\varepsilon_{11} - 2\varepsilon_{22}),$$
$$\hat{m}_{33} = (H + 2M)(1/2 - 3\varepsilon_{11} - \varepsilon_{22}),$$
$$\hat{m}_{44} = (H + 2M)(1/2 - 3\varepsilon_{22} - \varepsilon_{11}).$$

Within the considered order of magnitude, the warp-dependent part of free energy (39) is the homogeneous quadratic form in terms of partial derivatives of the torsion components.

The eigenvalues of matrix $\hat{M}$ are

(41) $\quad s_1 = 2M \cdot (1 - 4\varepsilon_{11} - 4\varepsilon_{22}), \quad s_2 = (2M + 2\Lambda) \cdot (1 - 4\varepsilon_{11} - 4\varepsilon_{22}),$

$$s_3 = \left(M + \frac{H}{2}\right)(1 - 4\varepsilon_{11} - 4\varepsilon_{22}) + \frac{1}{2}\sqrt{\Delta_1},\ s_4 = \left(M + \frac{H}{2}\right)(1 - 4\varepsilon_{11} - 4\varepsilon_{22}) - \frac{1}{2}\sqrt{\Delta_1},$$

$$\Delta_1 = [1 - 8(\varepsilon_{11} + \varepsilon_{22})](H^2 - 4MH + 4M^2) +$$
$$+ 4(\varepsilon_{11}^2 + \varepsilon_{22}^2)(-12MH + 5H^2 + 20M^2) + \varepsilon_{11}\varepsilon_{22}(96M^2 - 160HM + 24H^2)$$

The conditions of stability of the matrix $\hat{M}$ could be expressed in terms of its eigenvalues (Chapter 3, [15]):

(42) $\quad s_i(\varepsilon_{11}, \varepsilon_{22}) > 0, \quad i = 1,2,3,4.$

If all pre-strains vanish, the matrices coincide $\hat{M} \equiv M$. If one of the eigenvalues (41) turns to be negative, the medium destabilizes. As the strain remains constant, the appearance of new defects contributes to unlimited energy release and, therefore, instability of material. The state with non-vanishing torsion possesses less free energy density, as the defects-free state without torsion. The new defects arise in the medium and the plastic flow emerges. The plastic flow is treated in this study as the transition to the phase with non-vanishing torsion.

### 4.1 Isotropic tension

The conditions of positive definiteness of the quadratic form (39) in terms of applied homogeneous strains could be further simplified in the case of isotropic tension $\varepsilon_{11} = \varepsilon_{22} = \varepsilon$. The characteristic polynomial in this case is

$$F_1(x) = |\hat{M} - Ex| = (2M(1 - 8\varepsilon) - x)^2 (2(M + \Lambda)(1 - 8\varepsilon) - x)((1 - 8\varepsilon)H - x).$$

The conditions of positive definiteness of quadratic form in this case read

(43) $\quad s_{1,2} = 2M \cdot (1 - 8\varepsilon) > 0, s_3 = H \cdot (1 - 8\varepsilon) > 0, s_4 = (2M + 2\Lambda) \cdot (1 - 8\varepsilon) > 0.$

The conditions (43) demonstrate, that the medium with defects is stable under isotropic tension until the critical tensile strain attains its critical value

(44) $\quad \varepsilon < \varepsilon_{c.1} = 1/8.$

Remarkable, that the critical instability tensile strain is independent upon the material constants of material $M, \Lambda, H$.

### 4.2 Uniaxial tension

The case of uniaxial tension $\varepsilon_{11} = \varepsilon, \varepsilon_{22} = 0$ could be considered analogously. The characteristic polynomial reads

$$F_2(x) = |\hat{M} - Ex| = [2M(1 - 8\varepsilon) - x][2(M + \Lambda)(1 - 8\varepsilon) - x]$$
$$[(\varepsilon^2(4M^2 + H^2) - 2MH(14\varepsilon^2 - 8\varepsilon + 1)) + (2M + H)(1 - 4\varepsilon)x - x^2].$$

The conditions of positive definiteness of quadratic form in this case are the following

$$\text{(45)} \quad s_1 = 2\text{M}\cdot(1-8\varepsilon) > 0, \qquad s_2 = (2\text{M}+2\Lambda)\cdot(1-8\varepsilon) > 0,$$

$$s_{3,4} = \left(\text{M}+\frac{\text{H}}{2}\right)(1-4\varepsilon) \pm \frac{1}{2}\sqrt{\Delta_2} > 0,$$

$$\Delta_2 = 4\varepsilon^2(20\text{M}^2 - 12\text{HM} + 5\text{H}^2) + (\text{H}-2\text{M})^2(1-8\varepsilon)$$

The critical value of uniaxial tension is

$$\text{(46)} \quad \varepsilon_{2.c} = \min\left(\frac{1}{8}, \frac{\sqrt{2\text{MH}}(2\text{M}+\text{H})-8\text{HM}}{4\text{M}^2 - 28\text{MH} + \text{H}^2}\right).$$

When the uniaxial tension achieves its critical value $\varepsilon_{2.c}$, the spontaneous generation of defects occurs. The Fig.1 demonstrates the dependence of eigenvalues $s_i, i=1,..,4$ as the functions of uniaxial pre-strain $\varepsilon$. The instability and phase transition to plastic flow occurs when the lowest eigenvalue is zero.

### 4.3 Pure shear

The case of pure shear $\varepsilon_1 = -\varepsilon_2 = \gamma/2$ could be studied analogously.

$$F_3(x) = |\hat{\mathbf{M}} - \mathbf{E}x| = [2\text{M}-x][2(\text{M}+\Lambda)-x]\left[(\gamma^2(4\text{M}^2+\text{H}^2) - 2\text{MH}(2\gamma^2-1)) + (2\text{M}+\text{H})x - x^2\right]$$

The eigenvalues in this case are

$$\text{(47)} \quad s_1 = 2\text{M} > 0, \; s_2 = 2\text{M}+2\Lambda > 0,$$

$$\text{(48)} \quad s_{3,4} = \text{M}+\frac{\text{H}}{2} \pm \frac{1}{2}\sqrt{\Delta_3},$$

$$\Delta_3 = (\text{H}-2\text{M})^2 + 4(\text{H}+2\text{M})^2\gamma^2.$$

Every one of the eigenvalues $s_i(\gamma,-\gamma) > 0, \; i=1,2,3$ is positive for all values of $\gamma$. These conditions provide the upper bounds for stability of medium containing geometrical defects.
The eigenvalue $s_4$ could turn, however, to be negative for sufficiently high shear $\gamma$. The solution of the equation $s_4(\gamma,-\gamma) = 0$ is

$$\text{(49)} \quad \gamma_{c.2} = \frac{\sqrt{2\text{MH}}}{2\text{M}+\text{H}}.$$

The Fig.2 displays the eigenvalues $s_i, i=1,..,4$ as the functions of shear $\gamma$. The plastic flow occurs when the lowest eigenvalue vanishes. If condition $|\gamma| < \gamma_{c.2}$ is fulfilled, all four eigenvalues of matrix $\hat{\mathbf{M}}$ are positive. If $|\gamma| > \gamma_{c.2}$ the fourth eigenvalue will be negative and the instability in form of emergence of new topological defects occurs. In other words, the medium undergoes the spontaneous symmetry breaking in form of emerging dislocations and plastic flow.

## 5 Stability conditions and spontaneous symmetry breaking in three-dimensional case

The similar study could be performed in three-dimension case also. The Lagrangian of the elastic medium depends in three-dimensional case on 9 connection components:

$$\tilde{\mathbf{v}} = \|T_{11}, T_{12}, T_{13}, T_{21}, T_{22}, T_{23}, T_{31}, T_{32}, T_{33}\|^T.$$

The metric tensors in the reference state $s$ and deformed state $S$ are taken in the form

$$\text{(50)} \quad \mathbf{g} = \mathbf{G} = \begin{bmatrix} 1+2\varepsilon_{11}\cdot\delta & 1+2\varepsilon_{12}\cdot\delta & 1+2\varepsilon_{13}\cdot\delta \\ 1+2\varepsilon_{12}\cdot\delta & 1+2\varepsilon_{22}\cdot\delta & 1+2\varepsilon_{23}\cdot\delta \\ 1+2\varepsilon_{13}\cdot\delta & 1+2\varepsilon_{23}\cdot\delta & 1+2\varepsilon_{33}\cdot\delta \end{bmatrix}.$$

Note, that for brevity the strains assumed to be homogeneous and do not depend on coordinates. In three dimensional space the Ricci tensor reads

(51) $\quad \mathbf{R} = \mathbf{\Psi} \cdot \delta + (\mathbf{\Xi} + \overline{\mathbf{\Xi}}) \cdot \delta^2 + o(\delta^2)$

The components of tensors (48) in the orthogonal Cartesian coordinate system are the following:

$$\mathbf{\Psi} = \begin{bmatrix} \partial_2 T_{31} - \partial_3 T_{21} & \partial_3 T_{11} - \partial_1 T_{31} & \partial_1 T_{21} - \partial_2 T_{11} \\ \partial_3 T_{11} - \partial_1 T_{31} & \partial_3 T_{12} - \partial_1 T_{32} & \partial_1 T_{22} - \partial_1 T_{32} \\ \partial_1 T_{21} - \partial_2 T_{11} & \partial_1 T_{22} - \partial_1 T_{32} & \partial_1 T_{23} - \partial_2 T_{13} \end{bmatrix}$$

$$\mathbf{\Xi} = -2 \begin{bmatrix} \varepsilon_{22}\partial_2 T_{31} - \varepsilon_{33}\partial_3 T_{21} & \varepsilon_{33}\partial_3 T_{11} - \varepsilon_{11}\partial_1 T_{31} & \varepsilon_{11}\partial_1 T_{21} - \varepsilon_{22}\partial_2 T_{11} \\ \varepsilon_{33}\partial_3 T_{11} - \varepsilon_{11}\partial_1 T_{31} & \varepsilon_{33}\partial_3 T_{12} - \varepsilon_{11}\partial_1 T_{32} & \varepsilon_{11}\partial_1 T_{22} - \varepsilon_{22}\partial_2 T_{12} \\ \varepsilon_{11}\partial_1 T_{21} - \varepsilon_{22}\partial_2 T_{11} & \varepsilon_{11}\partial_1 T_{22} - \varepsilon_{22}\partial_2 T_{12} & \varepsilon_{11}\partial_1 T_{23} - \varepsilon_{22}\partial_2 T_{13} \end{bmatrix}$$

The components of tensor $\widetilde{\overline{\mathbf{\Xi}}}$ are the quadratic forms of torsion components:

$$\overline{\Xi}_{11} = -T_{33}T_{22} + \frac{1}{2}T_{11}T_{22} + \frac{1}{2}T_{11}T_{33} + T_{32}^2 + T_{23}^2 + \frac{1}{2}T_{33}^2 + \frac{1}{2}T_{22}^2,$$

$$\overline{\Xi}_{22} = -T_{33}T_{11} + \frac{1}{2}T_{33}T_{22} + \frac{1}{2}T_{11}T_{22} + T_{31}^2 + T_{13}^2 + \frac{1}{2}T_{33}^2 + \frac{1}{2}T_{11}^2,$$

$$\overline{\Xi}_{33} = -T_{22}T_{11} + \frac{1}{2}T_{22}T_{33} + \frac{1}{2}T_{33}T_{11} + T_{21}^2 + T_{12}^2 + \frac{1}{2}T_{11}^2 + \frac{1}{2}T_{22}^2,$$

$$\overline{\Xi}_{12} = \overline{\Xi}_{21} = T_{33}T_{21} - T_{32}T_{31} - T_{13}T_{23} + \frac{1}{2}T_{12}(T_{33} - T_{22} - T_{11}),$$

$$\overline{\Xi}_{13} = \overline{\Xi}_{31} = T_{22}T_{31} - T_{23}T_{21} - T_{12}T_{32} + \frac{1}{2}T_{13}(T_{22} - T_{11} - T_{33}),$$

$$\overline{\Xi}_{23} = \overline{\Xi}_{32} = T_{11}T_{32} - T_{31}T_{21} - T_{13}T_{12} + \frac{1}{2}T_{23}(T_{11} - T_{33} - T_{22}).$$

After the substitution of expression (48) in Lagrangian (21), the energy density reads

(52) $\quad L_\kappa = \frac{1}{2}\tilde{\mathbf{v}}^T \tilde{\mathbf{M}} \tilde{\mathbf{v}}$

Consider firstly the energy density for the strain-free elastic medium. If all pre-strains absent, the characteristic polynomial of matrix $\tilde{\mathbf{M}}$ is

$$F_4(x) = |\tilde{\mathbf{M}} - \mathbf{E}x| = (2H - x)^2(2M - x)^4(2x^2 - 3x(H + 2M) + 8MH)(4M + 6\Lambda - x)$$

The positive definiteness of the Lagrangian (49) could be expressed in terms of eigenvalues of matrix $\tilde{\mathbf{M}}$:

$$M > 0, \quad H > 0, \; 2M + 3\Lambda > 0,$$

(53) $\quad 3H + 6M + \sqrt{36M^2 - 28MH + 9H^2} > 0,$

$$3H + 6M - \sqrt{36M^2 - 28MH + 9H^2} > 0.$$

The conditions of positive definiteness of the quadratic form (49) in terms of applied homogeneous strains could be expressed easily in the case of isotropic tension $\varepsilon_{11} = \varepsilon_{22} = \varepsilon_{33} = \varepsilon$. The Lagrangian of pre-strained medium in this case is

(54) $\quad L_\kappa = \frac{1}{2}\tilde{\mathbf{v}}^T(1 - 8\varepsilon)\tilde{\mathbf{M}}\tilde{\mathbf{v}}.$

All eigenvalues of matrix $(1 - 8\varepsilon)\tilde{\mathbf{M}}$ vanish, if $1 - 8\varepsilon = 0$. The critical isotropic tension strain is therefore $\varepsilon < \varepsilon_{c.1} = 1/8$. The specified method could be applied for various homogeneous strain states, but the final expressions are bulky.

# 6    Conclusions

The classical physical interpretation of plastic flow is summarized in [16]. In this Article, the emergence of torsion and curvature – or, in language of physics, emergence of defects – is treated as the phase transition from elastic, defect-free state, equipped with Euclidean geometry to plastic, defect-rich, equipped with Riemann-Cartan geometry. We introduce the simple model of material with Lagrangian of defects, which is quadratic in components of Ricci tensor. The stability conditions for the strain-free medium impose certain conditions on the coefficients of quadratic form for free energy. The free energy in the pre-strained medium remains stable with respect to defect-emerging for considerably lower pre-strain and behaves as linear elastic medium. The stability conditions, however, could be violated, if the strain in material attains its critical value. In this case the instability in form of appearance of new topological defects – dislocations and disclinations - occurs. The medium undergoes the spontaneous symmetry breaking in form of emerging topological defects. Albeit the ultimate critical strain is significantly higher, than the observable inelastic strains, the model provide a straightforward explanation of the inelasticity effects as pure defect-induced, topological instability of medium. Consequently, the occurrence of plasticity and plastic flow treated as pure geometric instability in the proposed model. Remarkable, that the presented analysis requires no a priori knowledge about physical details of crystal structure or about interaction between atoms in medium.

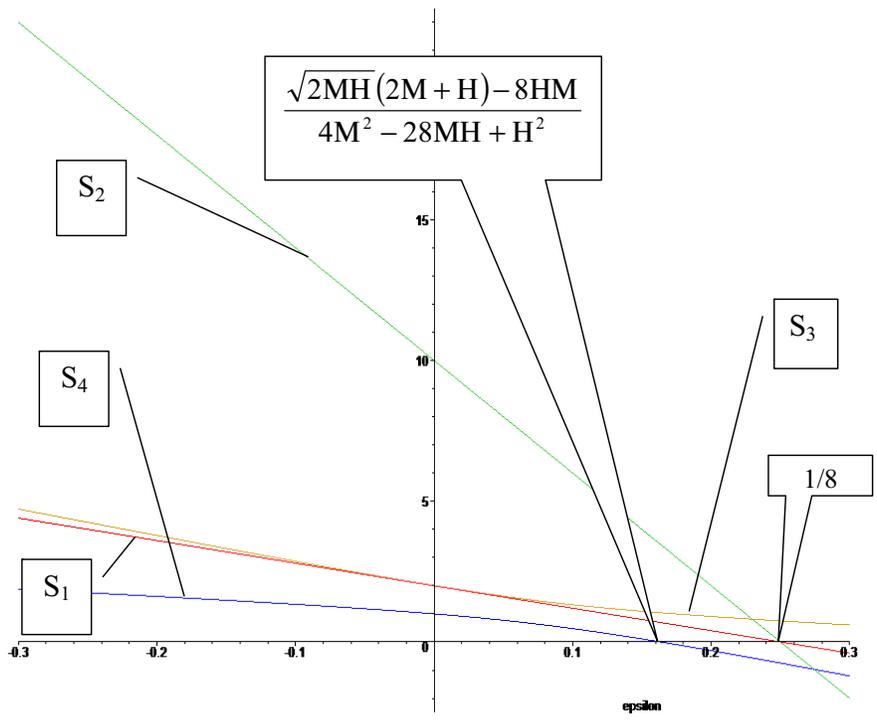

Fig. 1 The dependence of eigenvalues as function of uniaxial strain $\varepsilon$

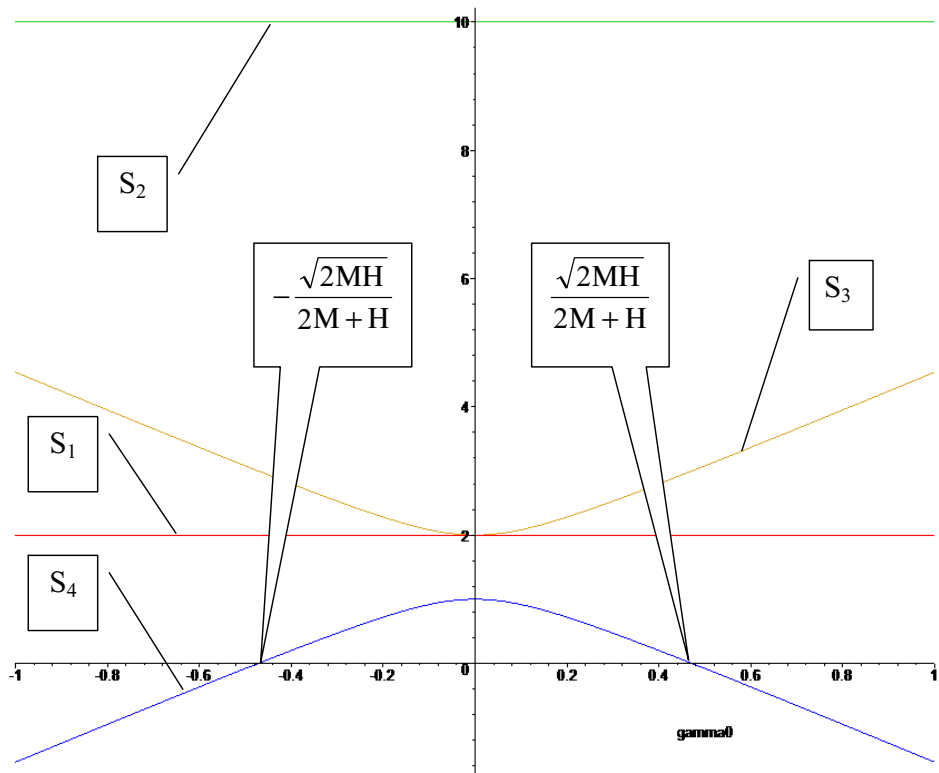

Fig.1 The dependence of eigenvalues as function of pure shear $\gamma$.